\documentclass[table]{article}
\usepackage{spconf,amsmath,graphicx,lipsum,amssymb}
\usepackage{booktabs,multirow}
\usepackage[dvipsnames]{xcolor}
\definecolor{best}{rgb}{0.8, 0.9, 1.0}
\definecolor{2ndbest}{rgb}{0.9, 1.0, 0.8}
\newcommand{\best}[1]{\colorbox{best}{\textbf{#1}}}
\newcommand{\secondbest}[1]{\colorbox{2ndbest}{#1}}
\usepackage{hyperref}
\usepackage{cite}

\def\x{{\mathbf x}}
\def\y{{\mathbf y}}
\def\E{{\mathbf E}}
\def\n{{\mathbf n}}
\def\btheta{{\boldsymbol \theta}}

\title{SPARSITY-DRIVEN PARALLEL IMAGING CONSISTENCY FOR IMPROVED SELF-SUPERVISED MRI RECONSTRUCTION}

\name{Ya\c{s}ar Utku Al\c{c}alar$^{*\dagger}$ \qquad Mehmet Ak\c{c}akaya$^{*\dagger}$}
\address{$^*$Department of Electrical \& Computer Engineering, University of Minnesota, MN, USA\\
$^\dagger$Center for Magnetic Resonance Research, University of Minnesota, MN, USA}

\begin{document}
\maketitle

\begin{abstract}
Physics-driven deep learning (PD-DL) models have proven to be a powerful approach for improved reconstruction of rapid MRI scans. In order to train these models in scenarios where fully-sampled reference data is unavailable, self-supervised learning has gained prominence. However, its application at high acceleration rates frequently introduces artifacts, compromising image fidelity. To mitigate this shortcoming, we propose a novel way to train PD-DL networks via carefully-designed perturbations. In particular, we enhance the k-space masking idea of conventional self-supervised learning with a novel consistency term that assesses the model’s ability to accurately predict the added perturbations in a sparse domain, leading to more reliable and artifact-free reconstructions. The results obtained from the fastMRI knee and brain datasets show that the proposed training strategy effectively reduces aliasing artifacts and mitigates noise amplification at high acceleration rates, outperforming state-of-the-art self-supervised methods both visually and quantitatively.

\begingroup
\renewcommand\thefootnote{}
\footnotetext{\textcopyright \textbf{ }2025 IEEE. Personal use of this material is permitted. Permission from IEEE must be obtained for all other uses, in any current or future media, including reprinting/republishing this material for advertising or promotional purposes, creating new collective works, for resale or redistribution to servers or lists, or reuse of any copyrighted component of this work in other works.}
\addtocounter{footnote}{-1}
\endgroup

\end{abstract}


\section{Introduction}
\label{sec:intro}
Magnetic resonance imaging (MRI) is a vital tool in modern radiology, but its long acquisition times pose challenges. Physics-driven deep learning (PD-DL) models have emerged as a powerful solution to accelerate MRI while preserving image quality~\cite{schlemper2018deep,hammernik2018VarNet,aggarwal2019MoDL,qin2019convolutional,yang2020admm_csnet}. Traditional supervised learning approaches require fully-sampled k-space data~\cite{schlemper2018deep,hammernik2018VarNet,aggarwal2019MoDL,mardani2018neural}, which is often unavailable or impossible to acquire in practical clinical settings. This has led to the exploration of alternative unsupervised methods, such as self-supervised learning~\cite{yaman2022mmssdu,chen2021equivariant} and generative modeling~\cite{jalal2021robust,chung2022scoreMRI}. Self-supervised learning, in particular, has gained attention as it alleviates the need for reference labels by leveraging partial k-space data to guide training~\cite{yaman2022mmssdu,zhang2024ccssdu} with theoretical guarantees~\cite{millard2023theoretical}. However, when applied at high acceleration rates, these methods may lead to residual artifacts, compromising reconstruction quality.

To tackle this challenge, we introduce \textbf{S}parsity-driven \textbf{P}arallel \textbf{I}maging \textbf{C}onsistent \textbf{S}elf-\textbf{S}upervised Learning via \textbf{D}ata \textbf{U}ndersampling (SPIC-SSDU), a method grounded in principles of parallel imaging MR reconstruction and sparsity-based image processing. SPIC-SSDU leverages carefully designed perturbations that avoid fold-overs in the field of view during acceleration, enforcing the network to accurately reconstruct these perturbations in alignment with clinical parallel imaging standards. The quality of these perturbations in the model output is then compared using a reweighted $\ell_1$ minimization principle to improve robustness and fidelity within the network. Our results on the fastMRI knee and brain datasets~\cite{knoll2020fastmri_dataset-journal-shorter} at acceleration rates of 6 and 8 demonstrate that the proposed framework surpasses state-of-the-art unsupervised and self-supervised methods, including multi-mask SSDU (MM-SSDU)~\cite{yaman2022mmssdu}, unsupervised learning from incomplete measurements (ULIM)~\cite{tachella2022ULIM}, and cycle-consistent SSDU (CC-SSDU)~\cite{zhang2024ccssdu}, both visually and quantitatively, while achieving artifact reduction comparable to supervised learning.

\begin{figure*}[t]
\centering
  \includegraphics[width=0.982\textwidth]{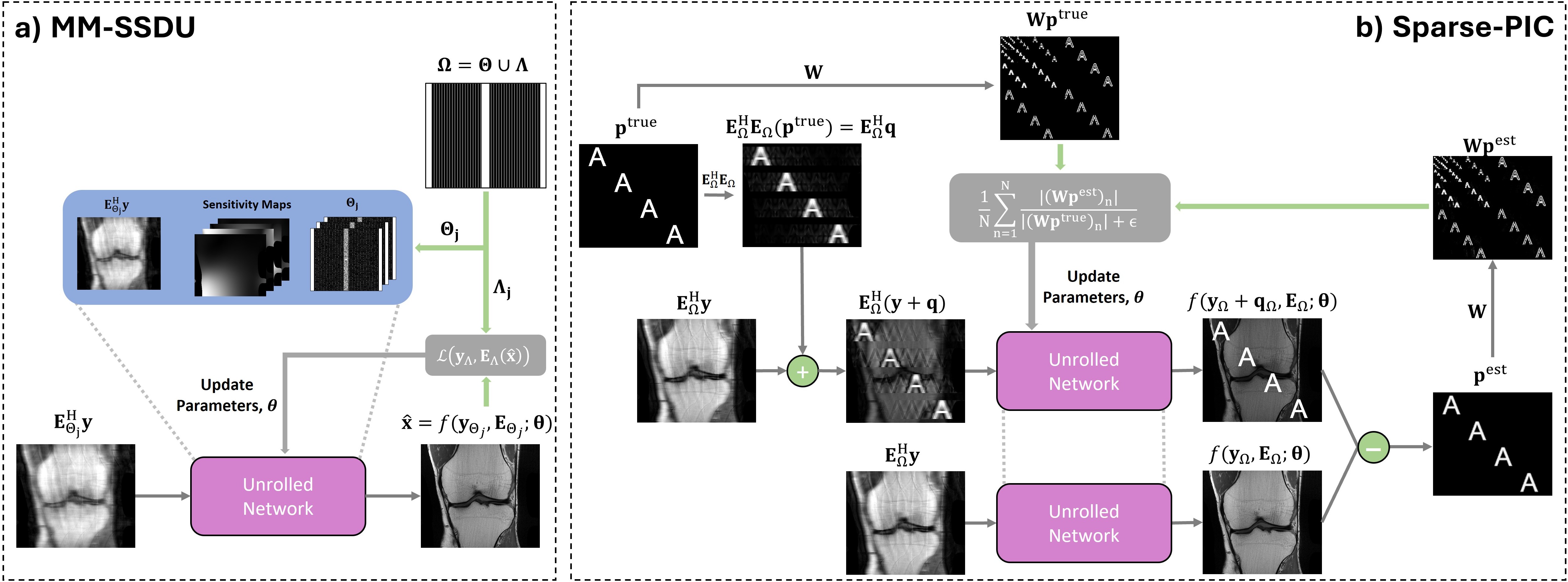}
  \caption{(a) Illustration of the MM-SSDU training process. The acquired measurements are  divided into two disjoint subsets: one used for data fidelity the PD-DL network and the other, unseen by the network, used to compute the loss. (b) Schematic representation of the Sparse-PIC (SPIC) block for a \emph{single} perturbation, which is designed to avoid aliasing overlaps in the phase encoding direction at a given acceleration rate. During the training, network outputs for perturbed and unperturbed inputs are compared to estimate the perturbation, and consistency is enforced $\ell_1$ weighting in a sparse domain to ensure the PD-DL model is consistent with clinical parallel imaging. In practice, multiple perturbations are applied, and their expectation is incorporated during training. Note that unlike MM-SSDU, all acquired measurements are utilized to create the network input.}
  \label{fig:methods}
\end{figure*}

\section{Background and Related Work}
\label{sec:related_work}

\subsection{MRI Inverse Problem and PD-DL Unrolling}
The MRI forward model that relates the underlying image \(\x\) to the acquired k-space data \(\y_\Omega\) is given as:
\begin{equation}
    \y_\Omega = \E_\Omega \x + \n,
\end{equation}
where $\Omega$ is the k-space undersampling pattern, $\E_\Omega$ is the multi-coil encoding operator that incorporates partial Fourier sampling, coil sensitivities, and the undersampling pattern, and $\n$ is measurement noise. Parallel imaging techniques leverage redundancies across coil sensitivities to reconstruct \(\x\) \cite{pruessmann1999sense}. Assuming i.i.d. Gaussian noise, the maximum likelihood estimate (MLE) of \(\x\) is given by:
\begin{equation}
    \x_{\text{PI}} = \arg\min_{\x} \|\y_\Omega - \E_\Omega \x\|_2^2 = (\E_\Omega^H \E_\Omega)^{-1} \E_\Omega^H \y_\Omega, \label{eq:par_img}
\end{equation}
where \(\E_\Omega^H\) denotes the Hermitian transpose of \(\E_\Omega\). This problem can be solved directly for certain $\Omega$ \cite{pruessmann1999sense}, or iteratively using conjugate-gradient (CG) in general \cite{pruessmann2001cgsense}. Alternatively, it can be addressed using k-space interpolation \cite{griswold2002grappa}. The inverse problem for MRI reconstruction is more generally formulated as a regularized least squares objective function:
\begin{equation}
    \arg \min_\x \| \y_\Omega - \E_\Omega \x \|_2^2 + \mathcal{R}(\x), \label{eq:mr_inv}
\end{equation}
where the first term ensures data fidelity with the measurements, and \( \mathcal{R}(\x) \) is a regularizer, whose corresponding proximal operator is learned implicitly by a neural network. PD-DL models typically unroll an iterative algorithm for solving \eqref{eq:mr_inv}~\cite{fessler2020SPM}, such as variable splitting~\cite{aggarwal2019MoDL}, for a fixed number of steps, where each step involves an alternating minimization problem. During supervised learning, the unrolled PD-DL network is trained in an end-to-end manner, where its output is compared to the fully-sampled reference data~\cite{aggarwal2019MoDL,hammernik2018VarNet}.

\begin{figure*}[t]
\centering
  \includegraphics[width=0.88\textwidth]{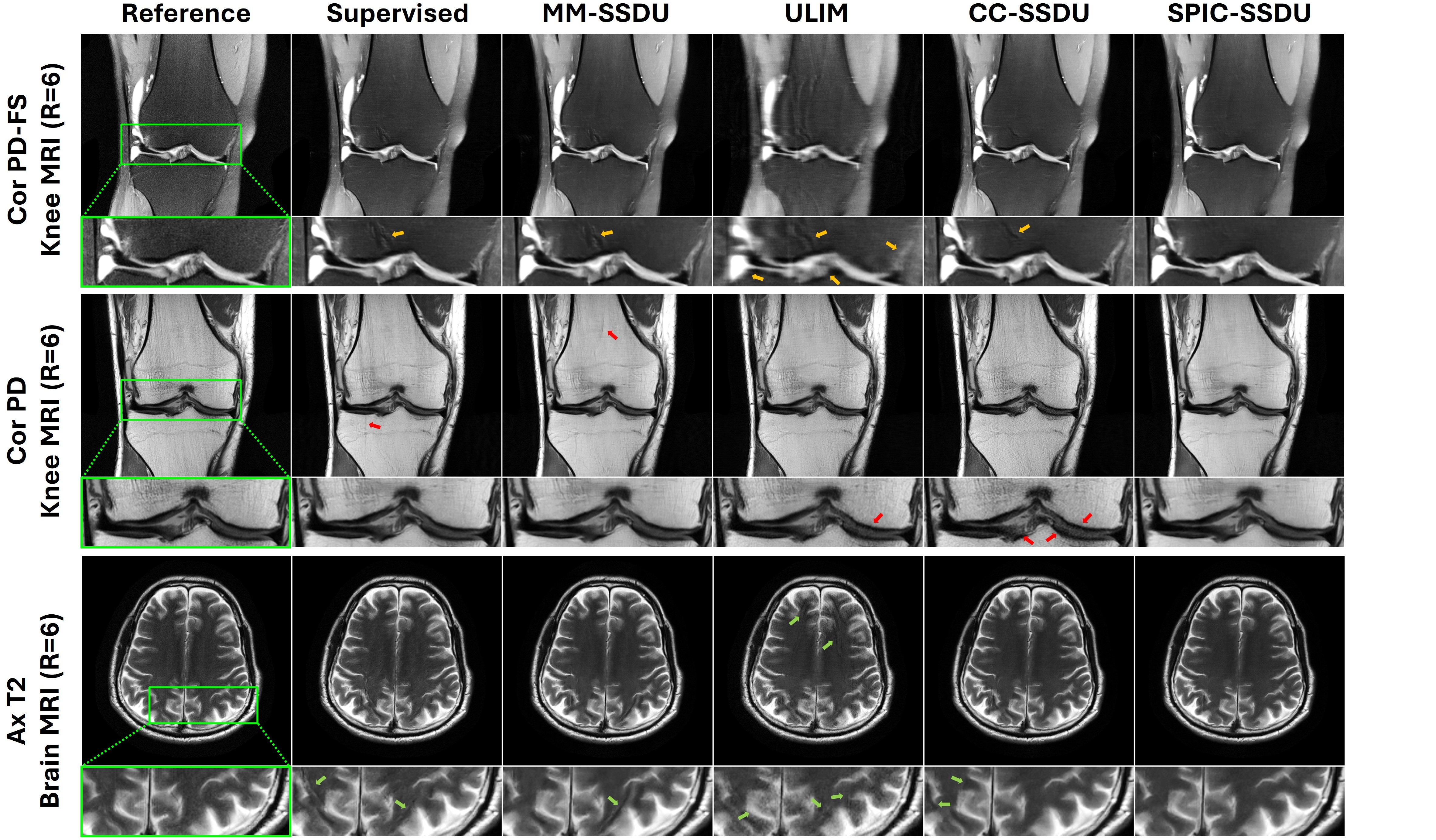}
  \caption{Representative slices reconstructed at $R = 6$ using equidistant undersampling from cor PD and cor PD-FS knee MRI, and ax T2 brain MRI. MM-SSDU exhibits noticeable artifacts in all datasets. ULIM shows good quality for cor PD but struggles with lower-SNR datasets like cor PD-FS and ax T2, where visible artifacts appear. CC-SSDU reduces artifacts in cor PD-FS and ax T2 but visibly amplifies noise in cor PD. SPIC-SSDU excels in artifact and noise reduction across all datasets, performing similar to supervised learning, while also mitigating residual artifacts seen in supervised methods.}
  \label{fig:R6}
\end{figure*}

\subsection{Self-Supervised Learning for MRI Reconstruction}
Obtaining fully-sampled data can be difficult or even impossible in various situations, primarily due to challenges such as organ movement, signal decay, and prolonged scan durations. To overcome this, self-supervised learning via data undersampling (SSDU) was proposed, where available measurement locations $\Omega$ are split into two disjoint subsets, $\Theta$ and $\Lambda$. $\Theta$ is used for data fidelity in the PD-DL network, while $\Lambda$ remains unseen by the network and is used to define the loss function. This method has been extended to multi-mask SSDU (MM-SSDU)~\cite{yaman2022mmssdu}, which employs multiple disjoint sets $\{(\Theta_k, \Omega_k)\}_{k=1}^K$, and the loss is computed as the expectation over these different subsets:
\begin{equation}
    \min_{\boldsymbol{\theta}} \; \mathbb{E}_{{\bf y}_{\Omega}}\Big\{\mathbb{E}_{(\Lambda,\Theta)} \big[ \mathcal{L}\left(\y_{\Lambda}, \E_{\Lambda}\left(f\left(\y_{\Theta}, \E_{\Theta}; \btheta \right)\right)\right) \big] \Big\}. \label{eq:mmssdu}
\end{equation}
A schmeatic of MM-SSDU algorithm is given in Fig.~\ref{fig:methods}a. Although it shows theoretical potential to match supervised learning performance~\cite{millard2023theoretical}, data splitting at high acceleration rates reduce training data substantially, leading to performance degradation~\cite{zhang2024ccssdu}.

A recent line of work to improve self-supervised reconstruction leveraged a cyclic measurement consistency approach~\cite{tachella2022ULIM,zhang2024ccssdu}, based on the idea that a well-trained PD-DL network should generalize effectively to new sampling patterns from similar distributions. To this end, let $\{\Delta_n\}$ be undersampling patterns matching $\Omega$'s distribution and acceleration rate $R$. As an example, for a uniform undersampling pattern $\Omega$, these can be shifted uniform patterns. Let the output of the PD-DL network be $\hat{\x}_\Omega = f\left( \y_{\Omega}, \E_{\Omega}; \btheta \right)$. Then, unsupervised learning from incomplete measurements (ULIM)~\cite{tachella2022ULIM} enforces cyclic consistency through the following minimization:
\begin{align}
    \min_{\btheta} \; \mathbb{E}_{{\bf y}_{\Omega}}&\Big\{\mathcal{L} \left( \y_{\Omega}, \E_{\Omega}\hat{\x}_\Omega \right)  \label{eq:ulim} \\
&+\beta \cdot \mathbb{E}_{\Delta} \big[ \mathcal{L} \left( \hat{\x}_\Omega, f\left( \E_{\Delta} \hat{\x}_\Omega, \E_{\Delta}; \btheta \right) \right) \big]\Big\}. \nonumber
\end{align}
Another cyclic-consistent approach, namely cycle-consistent SSDU (CC-SSDU)~\cite{zhang2024ccssdu}, takes this idea and implements a consistency $\Omega \to \Delta \to \Omega$ by focusing on minimizing the following:
\begin{align}
    \min_{\boldsymbol{\theta}} \; \mathbb{E}_{{\bf y}_{\Omega}}&\Big\{\mathbb{E}_{(\Lambda,\Theta)} \left[ \mathcal{L}\left(\y_{\Lambda}, \E_{\Lambda}\left(f\left(\y_{\Theta}, \E_{\Theta}; \btheta \right)\right)\right) \right]  \label{eq:cc_ssdu}\\
    &+\beta \cdot \mathbb{E}_{\Delta} \big[ \mathcal{L} \left( \y_{\Omega}, \E_{\Omega} f\left( \E_{\Delta} \hat{\x}_\Omega, \E_{\Delta}; \btheta \right) \big) \right]\Big\}. \nonumber
\end{align}
Although this approach shares similarities with ULIM's formulation in \eqref{eq:ulim}, it has several distinctions: In the first term, CC-SSDU uses MM-SSDU loss given in \eqref{eq:mmssdu} instead of enforcing similarity with all acquired measurements, which has found to have a better impact on PD-DL training~\cite{yaman2022mmssdu}. Furthermore, second term in \eqref{eq:cc_ssdu} enforces consistency only at the true acquired measurements instead of applying it across the full k-space as in \eqref{eq:ulim}.

\section{Self-Supervision via Sparsity-Driven Parallel Imaging Consistency}
\label{sec:methods}

\begin{figure*}[t]
\centering
  \includegraphics[width=0.88\textwidth]{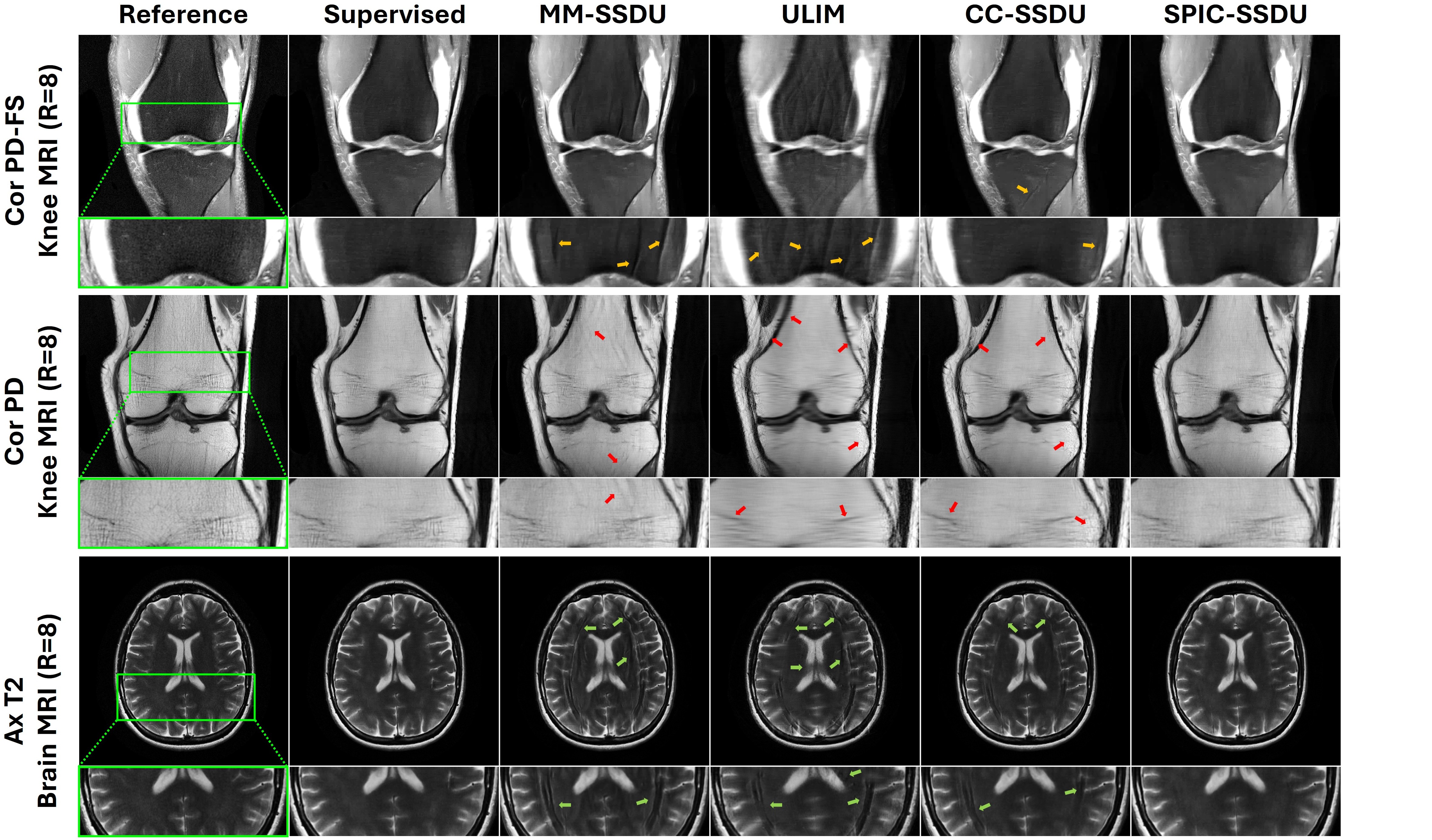}
  \caption{Illustrative slices reconstructed at $R=8$ using equidistant undersampling from cor PD, cor PD-FS knee MRI, and ax T2 brain MRI. MM-SSDU shows more pronounced artifacts compared to $R=6$. ULIM exhibits significant blurring in cor PD and amplifies artifacts in lower-SNR datasets like cor PD-FS and ax T2. CC-SSDU reduces artifacts in cor PD-FS and ax T2, though some are still visible. Both ULIM and CC-SSDU introduces high-frequency artifacts in cor PD. Proposed SPIC-SSDU outperforms all methods in artifact and noise reduction, achieving performance comparable to supervised learning.}
  \label{fig:R8}
\end{figure*}

In this work,  we take inspiration from classical MR reconstruction techniques, such as parallel imaging~\cite{pruessmann1999sense,griswold2002grappa} and compressed sensing~\cite{lustig2007sparse}, to introduce a novel self-supervised framework for training PD-DL networks, enabling high-quality reconstructions even at very high acceleration rates. Our key innovation involves augmenting the MM-SSDU loss, as defined in \eqref{eq:mmssdu}, with a novel consistency term, which we refer to as sparse parallel imaging consistency.

We achieve consistency between our network outputs and clinical parallel imaging reconstructions by introducing well-designed perturbations, $\{ {\bf p}_k \}$. For an acceleration rate of $R$, these perturbations are specifically structured to prevent aliasing artifacts from overlapping within the field of view. This design ensures that the resulting aliased perturbations can be resolved by \eqref{eq:par_img}. Specifically, for a given perturbation $\mathbf{p}$, let $\mathbf{q}_\Omega=\E_\Omega \mathbf{p}$. PD-DL models are expected to satisfy:
\begin{equation} \label{eq:perturb_approx}
    {\bf p} = f(\y_\Omega + \mathbf{q}_\Omega,\E_\Omega; \btheta) - f(\y_\Omega,\E_\Omega; \btheta),
\end{equation}
since under the assumption of well-designed perturbations that are recoverable by parallel imaging, PD-DL reconstructions should also recover the original perturbation. While $\ell_p$ norm based comparisons of both sides of \eqref{eq:perturb_approx} is a common approach, we propose to enforce this consistency within the sparse domain, as the perturbations themselves are sparse. This is achieved by a weighted $\ell_1$ norm with respect to the reference perturbation~\cite{candes2008enhancing} as:
\begin{equation}
    \mathcal{L}_{\text{s-pic}} = \mathbb{E}_{\bf p} \left[ \frac{1}{\mathrm{N}} \cdot \sum_{n=1}^\mathrm{N} \frac{|(\mathbf{W}{\bf p}^{\text{est}})_n|}{|(\mathbf{W}{\bf p}^{\text{true}})_n|+\epsilon} \right],
\end{equation}
where ${\bf p}^{\text{est}} \triangleq f(\y_\Omega + \mathbf{q}_\Omega,\E_\Omega; \btheta) - f(\y_\Omega,\E_\Omega; \btheta)$, $\mathbf{W}$ is the sparsifying transform, $N$ denotes the total number of sparse domain coefficients, and $\epsilon$ is a small number for numerical stability. Consequently, the final loss function for SPIC-SSDU is:
\begin{align}
    \min_{\boldsymbol{\theta}} \; \mathbb{E}_{{\bf y}_{\Omega}}&\Big\{\mathbb{E}_{(\Lambda,\Theta)} \big[ \mathcal{L}\left(\y_{\Lambda}, \E_{\Lambda}\left(f\left(\y_{\Theta}, \E_{\Theta}; \btheta \right)\right)\right) \big] \label{eq:spic_ssdu} \\
    &+ \beta \cdot \mathcal{L}_{\text{s-pic}}\Big\}, \nonumber
\end{align}
where $\beta$ denotes the trade-off parameter between the two loss terms. A schematic SPIC-SSDU training is given in Fig.~\ref{fig:methods}b.

\section{Evaluations}
\vspace{-.1cm}
\label{sec:evaluations}

\begin{table*}[t]
\small
\setlength{\tabcolsep}{4.2pt}
\caption{Quantitative results on coronal PD, coronal PD-FS, and axial T2-weighted datasets using equispaced undersampling patterns at $R=6$ and $R=8$. The \colorbox{best}{{\bf best}} and \colorbox{2ndbest}{second-best} values for \underline{self-supervised and unsupervised methods} are highlighted.}
\label{sample-table}
\begin{center}
\begin{tabular}{@{}p{2.7cm}cccccccccccc@{}}
\arrayrulecolor{black} \toprule
\multirow{4}{*}{Method} 
& \multicolumn{4}{c}{Cor PD, Knee MRI} 
& \multicolumn{4}{c}{Cor PD-FS, Knee MRI} 
& \multicolumn{4}{c}{Ax T2, Brain MRI} \\
\cmidrule(lr){2-5} \cmidrule(lr){6-9} \cmidrule(lr){10-13}
& \multicolumn{2}{c}{$R=6$} & \multicolumn{2}{c}{$R=8$} 
& \multicolumn{2}{c}{$R=6$} & \multicolumn{2}{c}{$R=8$} 
& \multicolumn{2}{c}{$R=6$} & \multicolumn{2}{c}{$R=8$} \\
\cmidrule(lr){2-3} \cmidrule(lr){4-5}
\cmidrule(lr){6-7} \cmidrule(lr){8-9}
\cmidrule(lr){10-11} \cmidrule(lr){12-13}
& PSNR$\uparrow$ & SSIM$\uparrow$ 
& PSNR$\uparrow$ & SSIM$\uparrow$ 
& PSNR$\uparrow$ & SSIM$\uparrow$ 
& PSNR$\uparrow$ & SSIM$\uparrow$ 
& PSNR$\uparrow$ & SSIM$\uparrow$ 
& PSNR$\uparrow$ & SSIM$\uparrow$ \\\midrule
Supervised~\cite{hammernik2018VarNet} & 39.07 & 0.952 & 36.15 & 0.921 & 34.45 & 0.824 & 33.12 & 0.795 & 35.42 & 0.921 & 33.59 & 0.907 \\
\arrayrulecolor{gray} \midrule
MM-SSDU~\cite{yaman2022mmssdu} & \secondbest{39.06} & \secondbest{0.950} & 35.19 & 0.905 & \secondbest{34.00} & 0.791 & \secondbest{32.51} & \secondbest{0.763} & \best{35.32} & 0.911 & 32.06 & 0.872 \\
ULIM~\cite{tachella2022ULIM} & 38.61 & 0.947 & 33.88 & 0.888 & 31.30 & 0.748 & 30.55 & 0.716 & 33.93 & 0.892 & 30.33 & 0.834 \\
CC-SSDU~\cite{zhang2024ccssdu} & 38.17 & 0.940 & \best{35.76} & \secondbest{0.912} & 33.98 & \secondbest{0.799} & 32.47 & 0.759 & 35.11 & \secondbest{0.912} & \secondbest{32.56} & \secondbest{0.881} \\
SPIC-SSDU \textbf{(Ours)}
& \best{39.10} & \best{0.951} & \secondbest{35.64} & \best{0.914} & \best{34.05} & \best{0.801} & \best{32.53} & \best{0.765} & \secondbest{35.24} & \best{0.917} & \best{32.74} & \best{0.887} \\
\arrayrulecolor{black} \bottomrule
\end{tabular}
\end{center}
\label{tab:quantitative}
\end{table*}

\subsection{Imaging Experiments and Implementation Details}
We conducted a comprehensive evaluation of our method with both qualitative and quantitative assessments. Fully-sampled multi-coil knee and brain MRI data from the New York University (NYU) fastMRI database~\cite{knoll2020fastmri_dataset-journal-shorter} were used. The knee dataset comprised of fully-sampled coronal proton density weighted (cor PD) and coronal proton density-weighted with fat suppression (cor PD-FS) images with a matrix size of 320$\times$368, whereas for brain MRI, axial T2-weighted (ax T2) images with a matrix size of 320$\times$320 were used. The knee and brain MRI datasets were collected using 15 and 16 receiver coils, respectively. Both datasets were retrospectively undersampled using a uniform/equidistant pattern with acceleration rates of $R=6$ and $8$ with 24 central k-space lines. Our focus was on equidistant sampling patterns that are clinically used, and which generate coherent artifacts that are substantially more difficult to remove compared to the incoherent artifacts resulting from random undersampling~\cite{knoll2019assessment}.

We unrolled a variable splitting algorithm~\cite{aggarwal2019MoDL,yaman2022mmssdu} to optimize \eqref{eq:mr_inv} for $T$=10 steps. Data fidelity units utilized CG with 15 iterations~\cite{aggarwal2019MoDL}. The regularizer was implemented as a CNN-based ResNet with 15 residual blocks~\cite{timofte2017ntire}. Each block consists of layers with 3×3 kernel sizes and 64 channels, leading to a total of 592,129 trainable parameters. The network is trained end-to-end with 300 slices from 10 subjects per dataset, and tested on 380 slices from 10 different knee MRI subjects and 300 slices from 10 distinct brain MRI subjects. Dual-tree complex wavelet transform (DTCWT)~\cite{selesnick2005dualtree-WT} was chosen as ${\bf W}$. 3 perturbations with 3 k-space masks were used and $\beta=5\cdot10^{-3}$ is selected as the trade-off parameter.

\begin{figure}[!b]
\centering
  \includegraphics[width=\columnwidth]{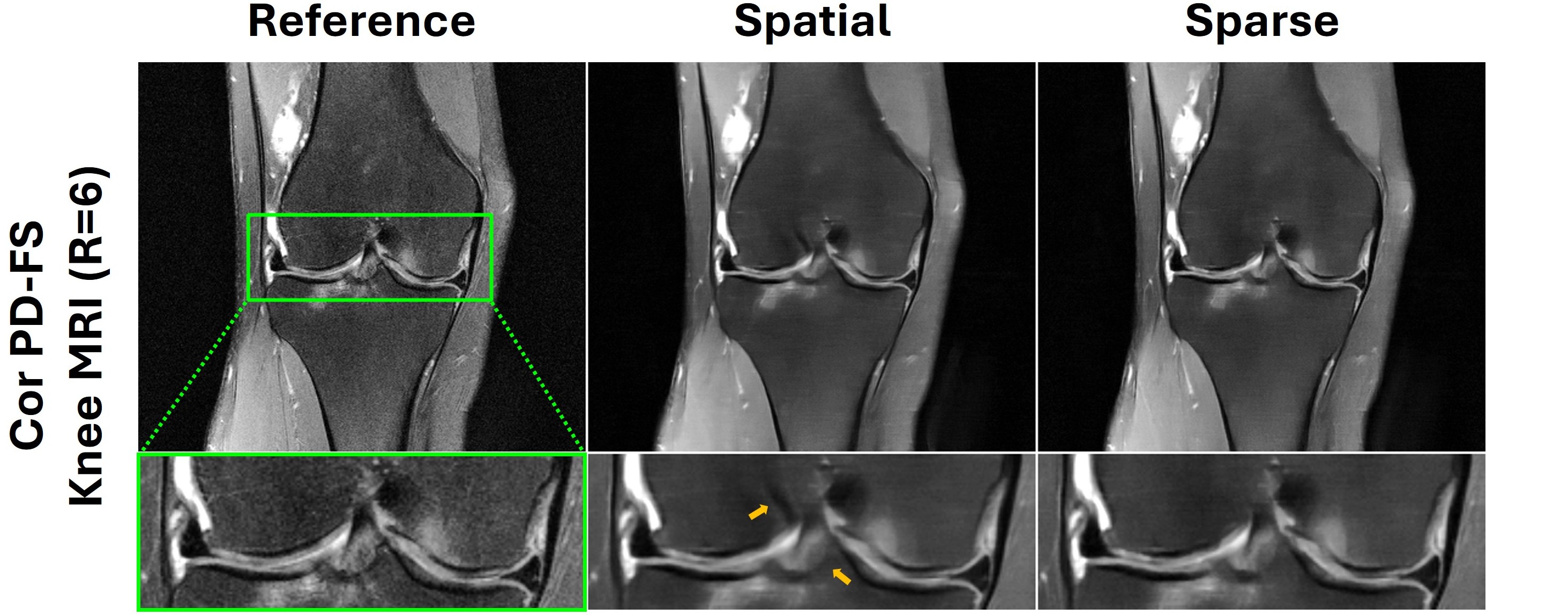}
  \caption{Effect of enforcing consistency between estimated and true perturbations in the sparse domain using the proposed weighted $\ell_1$ term compared to a standard normalized $\ell_2$ difference. The proposed processing in the sparse domain yields improved artifact reduction and sharper reconstructions.}
  \label{fig:ablation}
\end{figure}

\subsection{Comparison Methods}
We compared our method to supervised learning with a normalized $\ell_1$-$\ell_2$ loss~\cite{hammernik2018VarNet}, MM-SSDU~\cite{yaman2022mmssdu}, ULIM~\cite{tachella2022ULIM}, and CC-SSDU~\cite{zhang2024ccssdu}. ULIM and CC-SSDU were implemented using the formulations in \eqref{eq:ulim} and \eqref{eq:cc_ssdu}, respectively. For MM-SSDU, CC-SSDU, and SPIC-SSDU, $\rho=|\Lambda|/|\Theta|=0.4$ was selected as recommended in~\cite{yaman2022mmssdu}, and normalized $\ell_1$-$\ell_2$ loss was used as $\mathcal{L}$. To ensure fairness, the same PD-DL network with an equal number of parameters was used for all methods. Training times vary due to differences in the number of forward passes, but inference times are identical across methods and are the more critical factor in clinical settings. The results were quantitatively assessed using the structural similarity index (SSIM) and peak signal-to-noise ratio (PSNR).

\vspace{-.8mm}
\subsection{Results}
Fig.~\ref{fig:R6} and Fig.~\ref{fig:R8} show visual results on all datasets from $R=6$ and $R=8$ respectively. While MM-SSDU demonstrates slights artifacts at $R=6$, these artifacts become more pronounced at $R=8$. Additionally, while ULIM performs strongly on cor PD knee MRI at $R=6$, its robustness declines when applied to datasets with lower baseline SNR, such as cor PD-FS knee and ax T2 brain MRI, where visible artifacts emerge. At $R=8$, the limitations of ULIM become even more evident, with significant blurring observed in cor PD knee MRI and severe artifact amplification in cor PD-FS knee and ax T2 brain MRI. CC-SSDU has fewer artifacts at $R=6$ for cor PD-FS knee and ax T2 brain MRI, though some artifacts remain visible. However, at this acceleration rate, it exhibits visible noise amplification in cor PD. At $R=8$, CC-SSDU again shows fewer artifacts, although still visible, but begins to introduce high-frequency artifacts in cor PD that are also visible in ULIM.

SPIC-SSDU delivers the most effective artifact and noise reduction at both acceleration rates across all datasets, matching the performance of supervised learning, while also reducing artifacts present in supervised learning at $R=6$ across all datasets. The quantitative results presented in Tab.~\ref{tab:quantitative} support these visual observations, where SPIC-SSDU consistently outperforms other unsupervised and self-supervised methods in terms of PSNR and SSIM values.

Finally, we compared the performance of estimating added perturbations in the spatial domain by simply using an $\ell_2$ norm based loss $\mathcal{L_\text{pic}} = \frac{||{\bf p}^{\text{est}} - {\bf p}^{\text{true}}||_2}{||{\bf p}^{\text{true}}||_2}$ with the proposed sparse domain approach via \eqref{eq:spic_ssdu}, demonstrating that sparse domain estimation enables sharper, artifact-free reconstructions and superior overall quality (Fig.~\ref{fig:ablation}).

\section{Discussion and Conclusion}
\label{sec:discussion}
In this study, we introduce SPIC-SSDU, a novel framework that combines concepts from parallel imaging and compressed sensing to improve self-supervised PD-DL MRI reconstruction, particularly at high acceleration rates. By augmenting the MM-SSDU loss with a sparse parallel imaging consistency term, we ensure better alignment between network outputs and clinical reconstructions, reducing artifacts and noise. Our results show that SPIC-SSDU outperforms existing methods, such as MM-SSDU, ULIM, and CC-SSDU, in both artifact reduction and noise mitigation. We further note that while we only showed Cartesian uniform undersampling, the SPIC-SSDU framework can be applied to other patterns, including Cartesian random and radial undersampling, which were not the focus of this study.

\section{Acknowledgments}
This work was partially supported by NIH R01HL153146, NIH R01EB032830, NIH P41EB027061.

\fontsize{10}{11}\selectfont
\bibliographystyle{IEEEbib}
\bibliography{refs}

\end{document}